\documentclass{PoS}

\usepackage{amssymb}
\usepackage{fontenc}
\usepackage{times}
\usepackage{mathptmx}
\usepackage{graphicx}

\usepackage{amsmath}
\usepackage{subcaption}
\usepackage{multirow}

\title{Testing the $SU(4)$ degeneracy after low-mode removal with $J=2$ mesons}

\ShortTitle{Testing the $SU(4)$ degeneracy after low-mode removal with $J=2$ mesons}

\author{M.~Denissenya\\
        Institut f\"ur Physik, FB Theoretische Physik, Universit\"at Graz, Universit\"atsplatz 5,
         8010 Graz, Austria\\
        E-mail: \email{mikhail.denissenya@uni-graz.at}}

\author{L.~Glozman\\
        Institut f\"ur Physik, FB Theoretische Physik, Universit\"at Graz, Universit\"atsplatz 5,
         8010 Graz, Austria\\
         E-mail: \email{leonid.glozman@uni-graz.at}}

\author{\speaker{Markus Pak}\\
       Institut f\"ur Physik, FB Theoretische Physik, Universit\"at Graz, Universit\"atsplatz 5,
         8010 Graz, Austria\\
       E-mail: \email{markus.pak@uni-graz.at}}

\abstract{
Recent lattice results on Dirac low-mode removed $J=1,2$ mesons and $J=\frac{1}{2}$ baryons reveal the appearance of a new $SU(4)$ symmetry of confinement. Here the degeneracy of all $J=2$ iso-vector states within an irreducible representation of $SU(4)$ after low-mode removal is demonstrated. The $SU(4)$ symmetry contains $SU(2)_L \times SU(2)_R$ and $U(1)_A$ symmetries as subgroups and mixes all compontents $u_L,u_R,d_L,d_R$ of the two-flavor Dirac field. It implies the vanishing of the interaction of quarks with the color-magnetic field, which is shown by using the QCD Hamiltonian in Coulomb gauge.}

\FullConference{The 33rd International Symposium on Lattice Field Theory\\
		 14 -18 July  2015\\
		 Kobe International Conference Center, Kobe, Japan}

\begin{document}

\section{Motivation}
To understand the relation between confinement and chiral symmetry breaking, it is helpful to switch off one of these two 
non-perturbative phenomena of QCD and study its consequences. We switch off the chiral symmetry breaking effects on the valence quark sector by removing a small number of the lowest-lying Dirac eigenmodes from the valence quark propagators\footnote{Clearly, the gauge invariance and Lorentz invariance are not destroyed by this procedure. However, the quark field becomes slightly
nonlocal.} and determine the hadron masses\footnote{It should be kept in mind, that the sea quarks are not affected by this procedure.}~\cite{Glozman:2012fj}.

We observe, that this procedure does not destroy the confinement properties of hadrons. Bound states of quarks are still formed \cite{Denissenya:2014poa}.  The picture of hadrons acquiring their masses from chiral symmetry breaking only, i.e. via a dynamically generated quark mass, is no longer applicable. In contrast to all other hadrons, the $J=0$ ground state mesons disappear from the spectrum: The pseudoscalar particle $0^{-+}$ loses its role as a Goldstone-boson, Ref.~\cite{Denissenya:2014ywa} . 

The most interesting observation is, that after low-mode removal, the hadrons show a higher symmetry than the $SU(2)_L \times SU(2)_R \times U(1)_A$ symmetry of the QCD Lagrangian \cite{Denissenya:2014poa,Denissenya:2014ywa}. 
This symmetry has been identified from the degeneracies in the spectrum to be $SU(4)$, Refs.~\cite{Glozman:2014mka, Glozman:2015qva}.
Not only quark flavors of fixed chirality mix, but also the left- and right-handed components. It can be interpreted as the symmetry of confinement.  

It can be shown by using the QCD Hamiltonian in Coulomb gauge, that the interaction  of quarks with the color-magnetic field vanishes if the $SU(4)$ symmetry is manifest, Ref.~\cite{Glozman:2015qva}. The only interaction left is via the confining color-Coulomb interaction.  We refer to our system as the
\textit{dynamical QCD string}.  

Here we give a short overview of our recent results for $J=2$ mesons, which have been published in \cite{Denissenya:2015mqa}. 
Results on baryons can be found in Ref.~\cite{Denissenya:2015woa} and on $J=1$ mesons in Refs.~\cite{Denissenya:2014poa, Denissenya:2014ywa}. 
We summarize the physics implications, following the lines of arguments in \cite{Glozman:2015qva}.
 
\section{Lattice Setup}
\label{Chapter-Lattice-Setup}
The lattice we use is $16^3 \times 32$ and the lattice spacing is $a=0.118$ fm. 
The dynamical Overlap fermion gauge field configurations are provided by the JLQCD collaboration, Refs.~\cite{Aoki:2008tq,Aoki:2012pma}. 
The pion mass is $289(2)$ MeV, Ref.~\cite{Noaki:2008iy}. Our ensemble consists of $83$ gauge field configurations.  

The low-modes are removed from the quark propagators via the prescription:
\begin{align}
\label{low-mode-prescription}
S_{k}(x,y) = S_{\textsc{full}}(x,y) -  \sum_{i=1}^k \frac{1}{\lambda_i} v_i(x)  v^{\dagger}_i(y) \, ,
\end{align}
where $S_{\textsc{full}}(x,y)$ is the full propagator, $\lambda_i$ are the  eigenvalues of the overlap Dirac operator in a given gauge background 
and $v_i(x)$ the corresponding eigenvectors. The truncated propagator $S_{k}(x,y)$ depends on $k$, the number of low modes removed. 
We remove up to $k=30$ modes, which corresponds to an eigenvalue cutoff up to $225$ MeV. 

The Dirac operator is inverted on Gaussian and derivative based sources. The analysis of our hadron states is performed via the variational
method, see Refs.~\cite{Michael:1985ne,Luscher:1990ck,Blossier:2009kd}. 

\section{Low-mode removal on tensor mesons}
In Ref.~\cite{Denissenya:2015mqa} we have shown that the following degeneracies for the tensor meson iso-vector states
have to occur, if $SU(2)_L \times SU(2)_R \times U(1)_A$ is restored \footnote{The $a_2$ particle in $(1,0) \oplus (0,1)$ is denoted as 
$a'_2$ to distinguish it from the $a_2$ in $(1/2,1/2)_b$.}:
\begin{align}
\pi_2 \leftrightarrow a'_2  \; , \text{and} \; , 
a_2 \leftrightarrow \rho_2 \; . 
\end{align}
If $SU(4)$ is restored \textit{all} iso-vector states have to become mass degenerate:
\begin{align}
\pi_2 \leftrightarrow a'_2 \leftrightarrow a_2 \leftrightarrow \rho_2  \; .
\end{align}

\begin{figure}[t]
\centering
\includegraphics[angle=0,width=.48\linewidth]{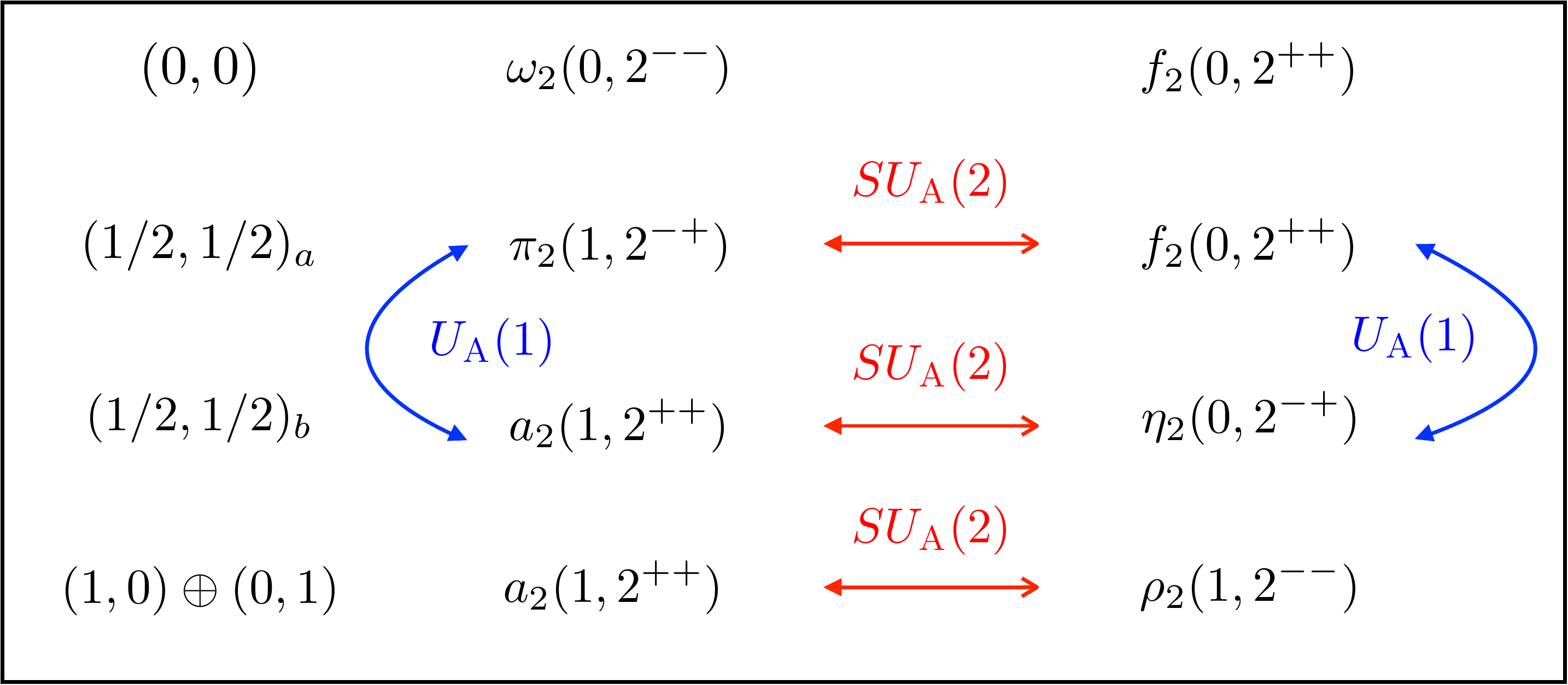}
\includegraphics[angle=0,width=.50\linewidth]{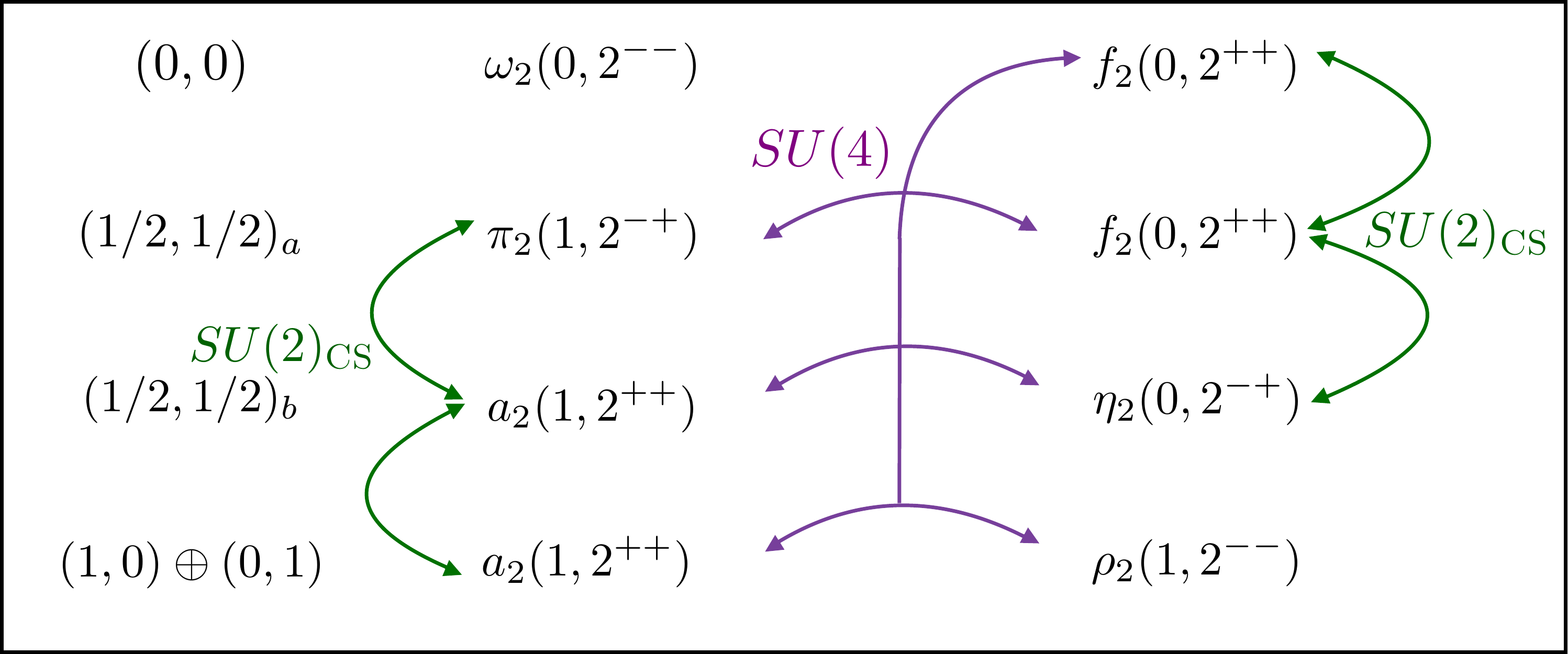}
\caption[Chiral-Parity group 1]{\sl \textit{Left:} Classification of states in the $SU(2)_L \times SU(2)_R
\times C_i$ group. $SU(2)_{\textsc{A}}$ is denoted by a red line and $U(1)_{\textsc{A}}$ by a blue line. If both these symmetries are restored, not all isovectors are mass degenerate. \textit{Right:} The $SU(4)$ 15-plet is denoted by a purple line. The singlet is the iso-scalar $\omega_2$. $SU(2)_{CS}$ is a subgroup of $SU(4)$ and connects the states with same isospin in distinct chiral representations. Figures are taken from \cite{Denissenya:2015mqa}.}
 \label{Table1}
\end{figure}

In Fig.~\ref{Table1} we compare these two different degeneracy patterns.  The figures are taken from Ref.~\cite{Denissenya:2015mqa}. 

  \begin{figure*}[htb]
    \centering 
      \begin{subfigure}[b]{0.45\textwidth}
        \centering
        \includegraphics[scale=0.55]{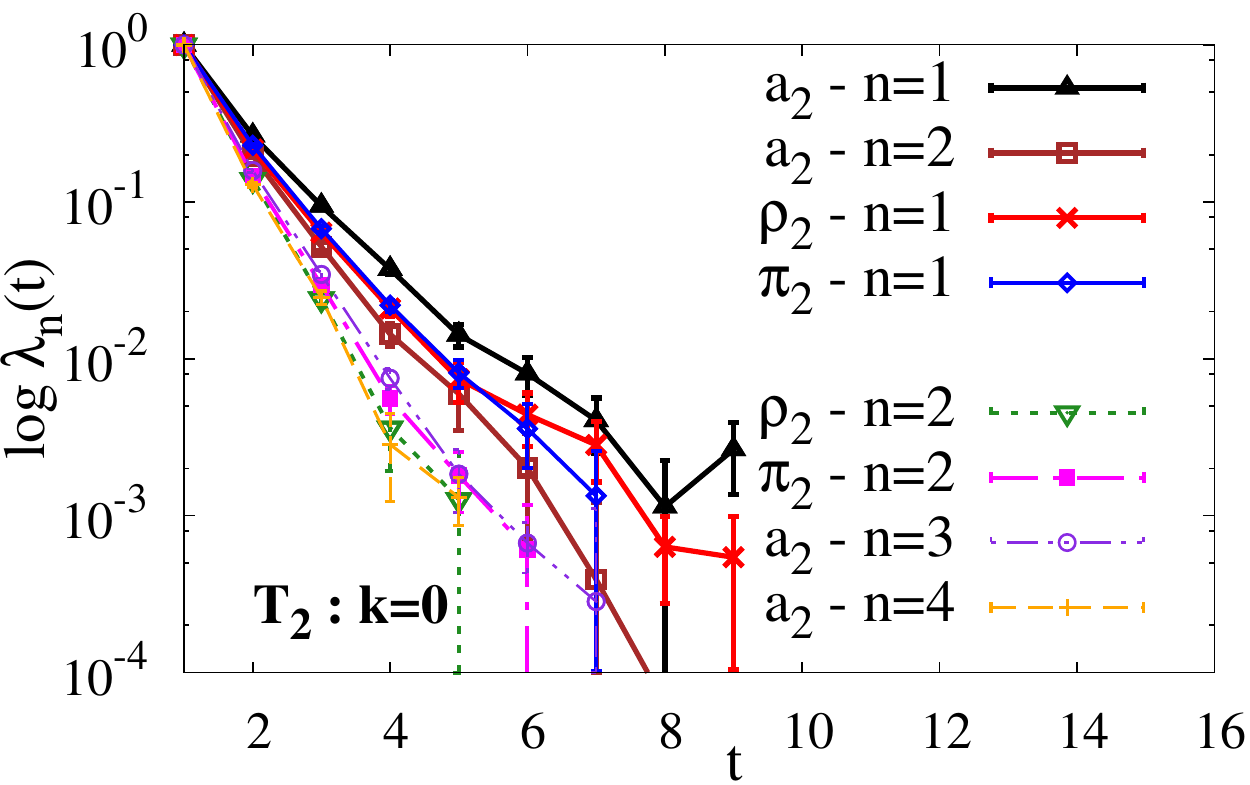}
        \caption{}
      \end{subfigure}
      % \hspace*{-12pt}
      \begin{subfigure}[b]{0.45\textwidth}
          \centering
          \includegraphics[scale=0.55]{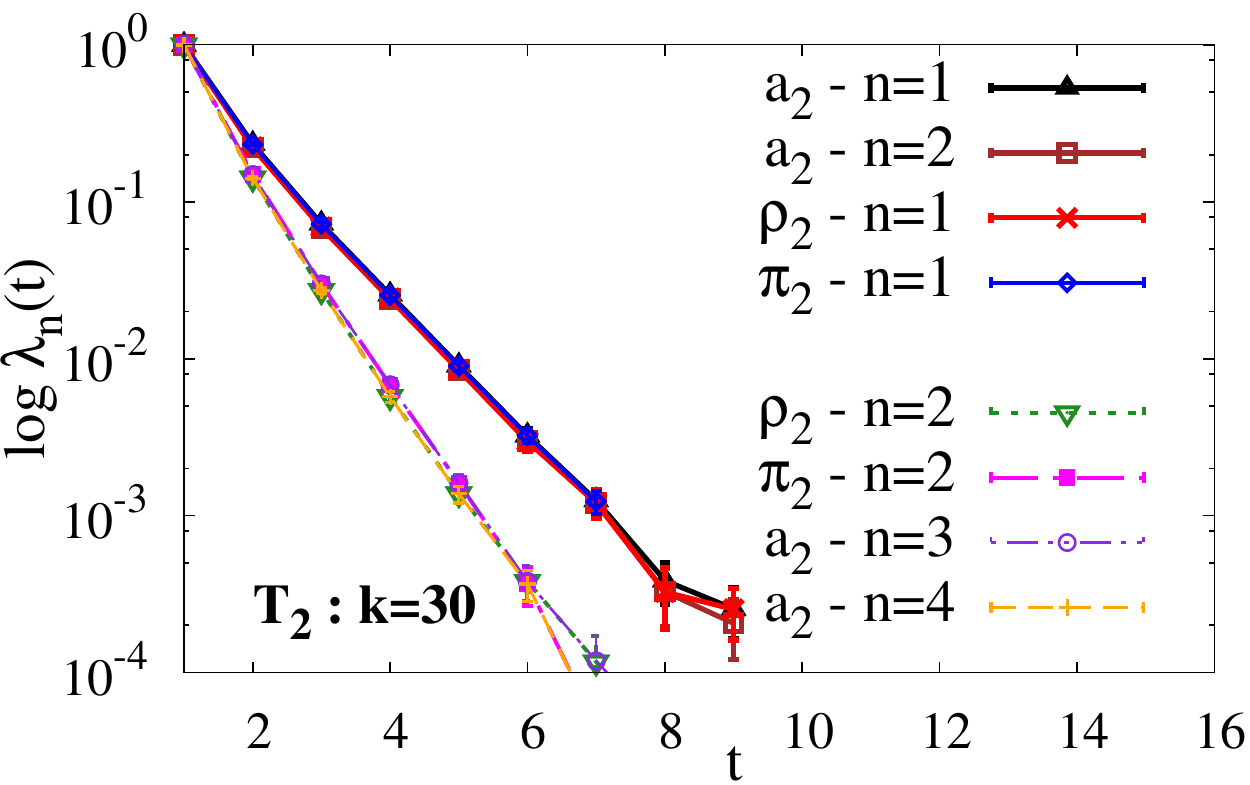}
          \caption{}
      \end{subfigure}
     \hspace*{24pt} \hfill\\
      \caption{Eigenvalues of the correlation matrix for  $J=2$ tensor mesons: (a) full case ($k=0$), (b) excluding $k=30$
low-lying modes. The label $T_2$ refers to the irreducible representation of the hypercubic group $O_{\text{h}}$. 
 Figures are taken from \cite{Denissenya:2015mqa}.
}\label{fig:j2T2ev}
   \end{figure*}

Via our lattice study, we can now identify, which of these symmetries are visible in the spectrum. It is sufficient to study iso-vectors in order to reveal if the $SU(4)$ symmetry occurs after low-mode removal or not. 

In Fig.~\ref{fig:j2T2ev} we compare the eigenvalues of the correlation matrix for the untruncated and low-mode truncated ($k=30$)
cases. We observe, that after low-mode removal \textit{all} correlators fall on the same curve, signalling
$SU(4)$ symmetry restoration. 

\begin{figure*}[htb]
    \centering 
      \begin{subfigure}[b]{0.45\textwidth}
        \centering
        \includegraphics[scale=0.50]{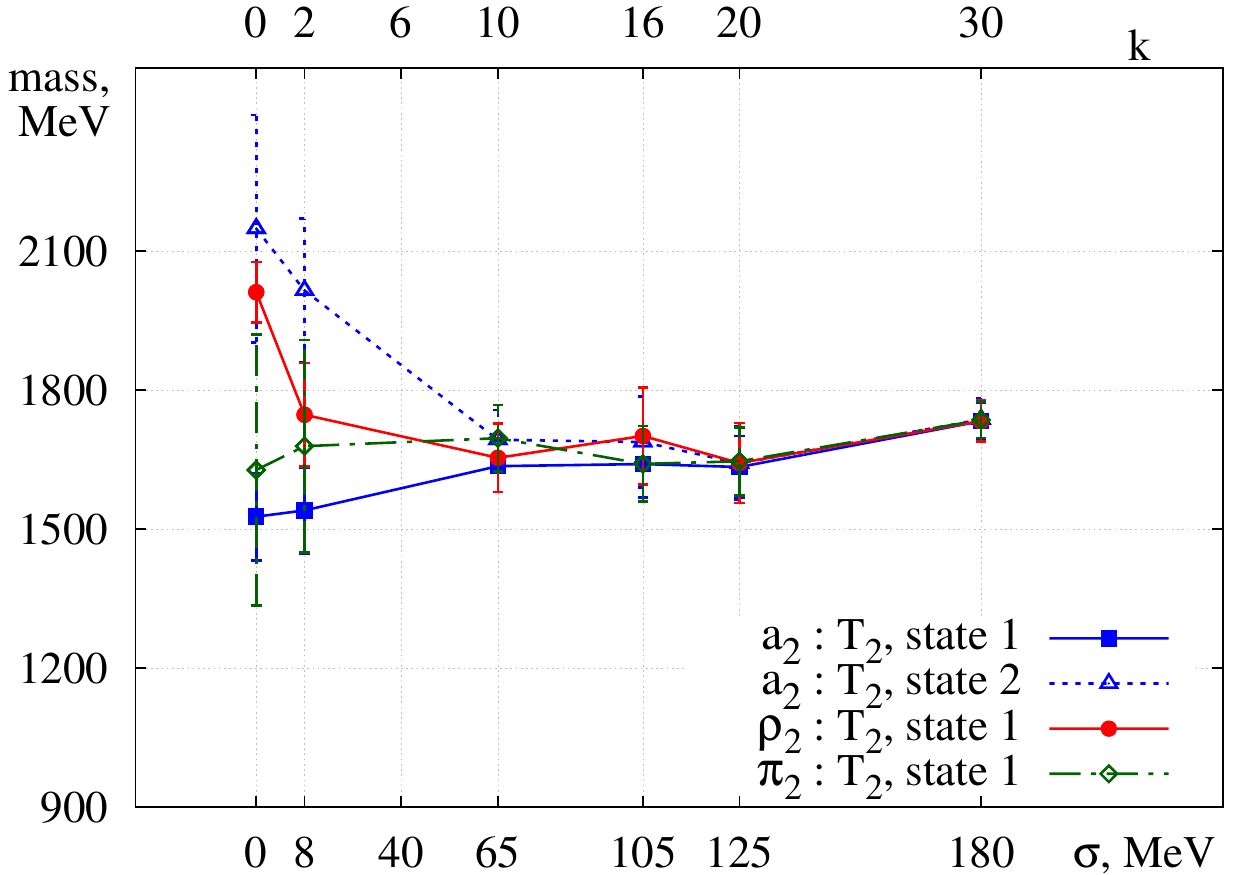}
        \caption{} \label{fig:a}
      \end{subfigure}
       \hspace*{-8pt}
      \begin{subfigure}[b]{0.45\textwidth}
          \centering
          \includegraphics[scale=0.50]{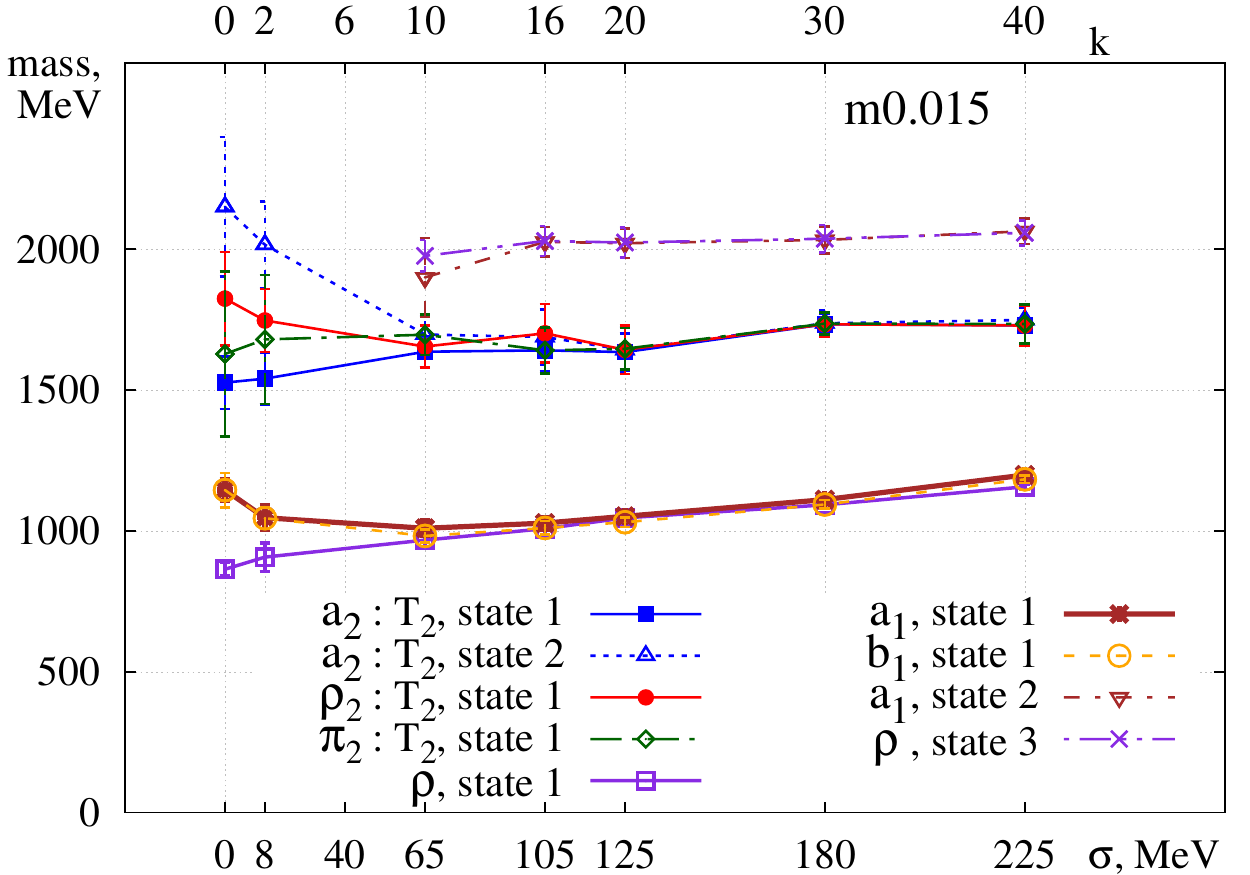}
          \caption{}\label{fig:b}
      \end{subfigure}
     \hspace*{24pt} \hfill\\
      
      \caption{Ground state and excited state masses of (a) $J=2$ mesons , (b) $J=1$ and $J=2$ mesons, with increasing number of truncation step $k$. The value $\sigma$ denotes the energy gap. Figures are taken from \cite{Denissenya:2015mqa}.
}\label{fig:j1j2T2}
   \end{figure*}

In Fig.~\ref{fig:a} we plot the masses for increasing number of truncation. After $k=20$ modes removed, which corresponds to an energy cutoff of $125$ MeV, the $SU(4)$ symmetry is visible in the spectrum. We emphasize, that the mass plateaus improve after the low-modes are removed
from the quark propagators. 

In Fig.~\ref{fig:b} we show the meson masses as a function of the truncation number $k$ for both $J=1$ and $J=2$ states. 
For the $J=1$ case the excited states are shown as well, beginning with truncation $k=10$, where it is possible to extract a state.
The energy levels of $J=1$ and $J=2$ mesons after removing the low-lying modes are clearly split. There is no indication of a higher symmetry than $SU(4)$, which would connect states of different spin $J$. However, one needs to take into account volume corrections to make a precise statement. Finite volume effects are currently analyzed using quenched gauge field configurations.

\subsection{Consequences of the $SU(4)$ symmetry} 
We have demonstrated that all isovector tensor mesons become degenerate after 
low-mode removal. We now stress the physics implications of it. We 
follow the lines of argument given in Ref.~\cite{Glozman:2015qva}. 

We use the QCD Hamiltonian in Coulomb gauge. We concentrate on the parts which couple quarks
to gluons:
\begin{align}
H_C &= \frac{g^2}{2} \int d^3 x \int d^3 y \; \mathcal{J}^{-1}[A] \rho^a(\boldsymbol{x}) \mathcal{J}[A]
F^{ab}(\boldsymbol{x}, \boldsymbol{y}) \rho^b(\boldsymbol{y})\; , \\ 
\label{HT}
H_T &= - g \int d^3 x \; \psi^{\dagger}(\boldsymbol{x}) \boldsymbol{\alpha} \cdot \boldsymbol{A} \psi(\boldsymbol{x}) \; . 
\end{align}

The first part $H_C$, referred to as Coulomb-interaction, because in the weak
coupling regime it comprises the color-Coulomb potential. It describes the interaction of non-abelian  color charge densities of quarks and gluons,
\begin{align}
\rho^a = \psi^{\dagger} T^a \psi - f^{a b c} A_i^b E_i^c \; , 
\end{align}
through the non-abelian Coulomb kernel $F^{ab}$. Here $f^{a b c}$ are the structure constants of color $SU(3)$ and $E_i$ is the color-electric field. The quantity $\mathcal{J}[A]$ is the determinant of the Faddeev-Popov operator.  
$H_C$ is invariant with respect to $SU(4)$ transformations. 
This part of the QCD Hamiltonian is confining, because it represents the
interaction between the color charges.

The second part, $H_T$, describes the interaction of quarks with
transverse gluons (i.e. with the color-magnetic field). It is not a $SU(4)$-singlet
and consequently its expectation value must vanish in a $SU(4)$-symmetric hadron wave function.
Therefore, it is only the interaction with the color electric-fields, which persists in an $SU(4)$-symmetric hadron after the near-zero mode truncation.   

\section{Summary and Conclusions}
\label{Conclusions}
Spin-2 iso-vector mesons have been analyzed with respect to the low-mode
removal of Dirac operator. 
It is found that after excluding around $20$ modes, the $SU(4)$ symmetry is visible in the spectrum.  
These results are fully in line with the $J=1$ case.  For the case of baryons \cite{Denissenya:2015woa} we have shown recently, that the $SU(4)$ symmetry is applicable as well. 

The $SU(4)$ emerges as a symmetry of hadrons after removing the low-lying modes from the quark propagators. It rotates both quark flavors and both the left- and right-handed fields into each other. It is shown \cite{Glozman:2015qva}, that after removing the low-modes, the only interquark
interaction left in the system is via the color-electric field.

\acknowledgments
The authors are thankful to  JLQCD collaboration for supplying us with
the overlap gauge configurations.
This work is supported by the Austrian Science Fund (FWF)
through the grant P26627-N27. Computations were carried out on clusters 
at ZID at the University of Graz and 
at the Graz University of Technology.

\end{document}